# INTEGRAL monitoring of unusually long X-ray bursts

**Jérôme Chenevez**[1]
*DTU Space – National Space Institute, Juliane Maries vej 30, 2100 Copenhagen, Denmark*
*E-mail:* jerome@space.dtu.dk

**Maurizio Falanga**
*CEA Saclay, DSM/DAPNIA, Service d'Astrophysique (CNRS FRE 2591), 91191 Gif-sur-Yvette, France*
*E-mail:* mfalanga@cea.fr

**Erik Kuulkers**
*ISOC, ESA/ESAC, Urb. Villafranca del Castillo, PO Box 50727, 28080 Madrid, Spain*
*E-mail:* Erik.Kuulkers@esa.int

**Søren Brandt**
*DTU Space – National Space Institute, Juliane Maries vej 30, 2100 Copenhagen, Denmark*
*E-mail:* sb@space.dtu.dk

**Niels Lund**
*DTU Space – National Space Institute, Juliane Maries vej 30, 2100 Copenhagen, Denmark*
*E-mail:* nl@space.dtu.dk

**Andrew Cumming**
*Physics Department, Mc Gill University, 3600 rue University, Montréal QC, H3A 2T8, Canada*
*E-mail:* cumming@physics.mcgill.ca

X-ray bursts are thermonuclear explosions on the surface of accreting neutron stars in low mass X-ray binaries. As most of the known X-ray bursters are frequently observed by INTEGRAL, an international collaboration have been taking advantage of its instrumentation to specifically monitor the occurrence of exceptional burst events lasting more than ~10 minutes. Half of the so-called intermediate long bursts registered so far have been observed by INTEGRAL. The goal is to derive a comprehensive picture of the relationship between the nuclear ignition processes and the accretion states of the system leading up to such long bursts. Depending on the composition of the accreted material, these bursts may be explained by either the unstable burning of a large pile of mixed hydrogen and helium, or the ignition of a thick pure helium layer. Intermediate long bursts are particularly expected to occur at very low accretion rates and make possible to study the transition from a hydrogen-rich bursting regime to a pure helium regime.

[1] Speaker

# 1.Introduction

## 1.1 Science background

X-ray bursters consist of a neutron star in a binary system with a low-mass companion star. The stars are sufficiently close that gas (largely hydrogen and helium) from the companion is accreted via its Roche lobe to the neutron star, accumulating on the surface, and forming a layer of hydrogen, which, due to the high temperature and density, soon transforms into a subsurface layer of helium. In the prevailing conditions on most accreting neutron stars, the nuclear fusion processes are stable and the emission from the neutron star surface is practically negligible compared to the emission from the accretion disk. However, for some specific accumulation rates, the burning layer reaches a critical thickness above which new unstable chains of nuclear reactions open up leading to thermonuclear runaways. Such thermonuclear flashes, called type I X-ray bursts, are characterized by a black-body emission with temperature above 20 million degrees ($kT \approx 2$ keV), and generally display a fast rise time followed by an exponential decay phase. This explosive thermonuclear burning can be recurrent and the accumulated gas is ignited every few hours to days, depending upon the accretion rate. The observation of such bursts from an X-ray binary system is a convenient way to identify the nature of the source. Indeed, only neutron stars (in opposition to black holes which have no surface) in low-mass binary systems present the necessary characteristics to produce X-ray bursts. However, only about a third of the known accreting neutron stars are X-ray bursters, because the bursting conditions imply a delicate balance between the accretion rate and the strength of the neutron star's magnetic field. Indeed, only relatively old neutron stars have a weak enough magnetic field to allow the accretion flow to cover a large enough area on the neutron star surface (for a review see Lewin et al., 1993; Cumming, 2004).

More than thirty years of observations of type I X-ray bursts have revealed a surprisingly rich spectrum of behaviour and have made it possible to investigate the nuclear processing on the surface of neutron stars, leading to a better understanding of their interior thermal structure, magnetic field, and spin. The instabilities of nuclear burning in relation with the X-ray bursts have received much theoretical attention, making clear that much of the accumulated hydrogen and helium is burned to heavy elements during the bursts (see, e.g., Strohmayer & Bildsten, 2006 for a recent review). Observations of X-ray bursts have in a few cases confirmed theoretical models of thermonuclear ignition and burning mechanisms. However, the behaviour of a majority of systems cannot be fully reconciled with theoretical predictions, suggesting there is additional physics at work.

One major observational characteristic of the X-ray bursts is their duration, from which three distinct classes are obvious (see Fig. 1): 1) bursts with durations up to a few tens of seconds represent 99% of the registered population, 2) about 15 events lasting several hours – and thus called superbursts – have been observed, and 3) a dozen of so-called intermediate long bursts have exhibited decay times of a few tens of minutes. The shortest bursts are consistent with pure helium flashes, while the presence of hydrogen lengthens the burst duration somewhat due to slow weak interactions involved in the hydrogen burning. At the longest end of the first class, bursts with durations of a few minutes are explained by the burning of a relatively large proportion of hydrogen. The latter two groups are the object of the present work, and are described in more details in the following.

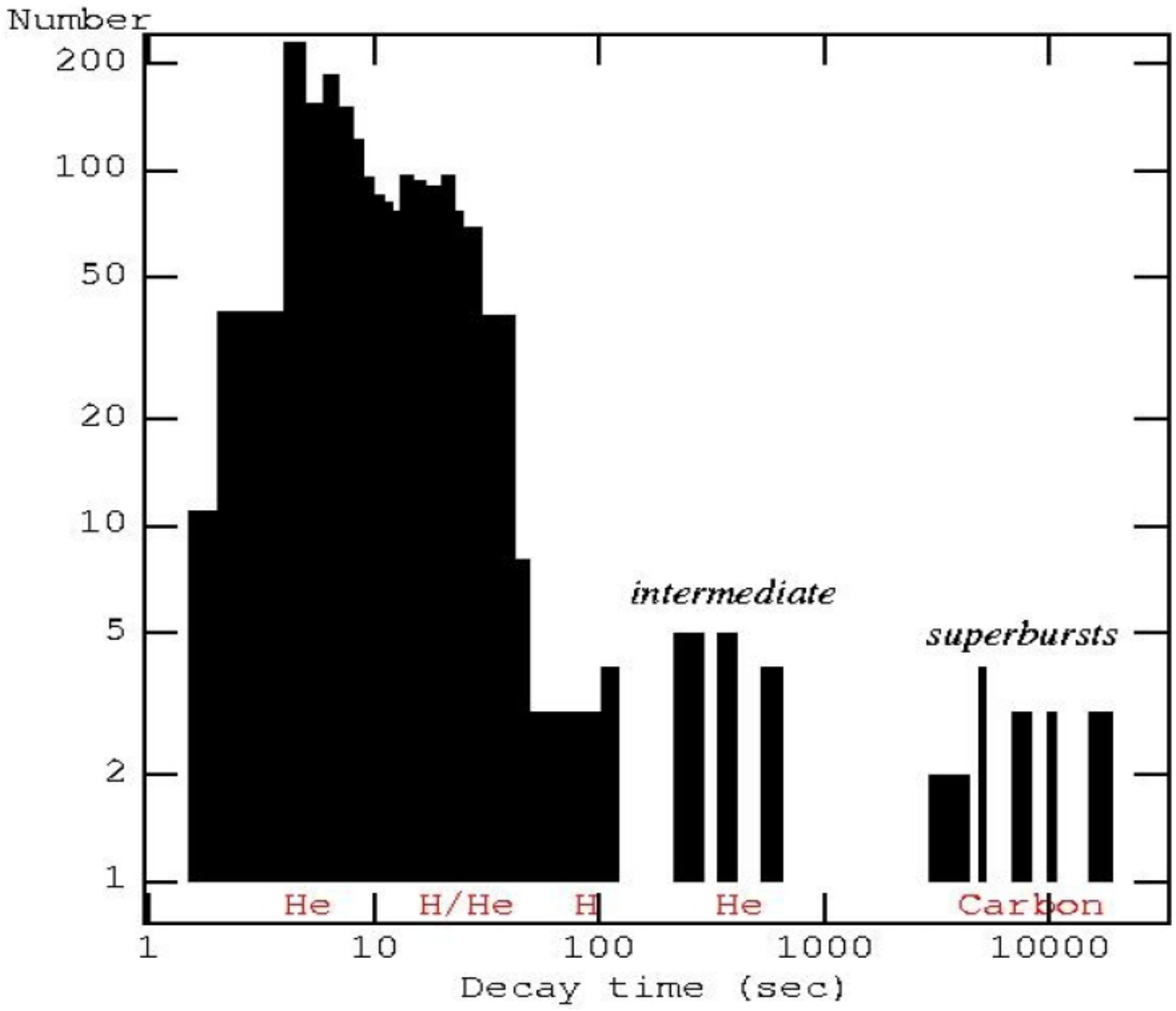

**Figure 1**: X-ray burst distribution as a function of their exponential decay times. The bursts up to 100s are a compilation of the RXTE sample (Galloway et al., 2008). Corresponding main burning fuel is indicated at the bottom.

### 1.1.1 Superbursts

Within the last 10 years a new type of X-ray bursts have been found (e.g., Cornelisse et al., 2000, 2002; Kuulkers et al., 2002) that last several hours and release about 1000 times more energy than usual type I bursts (Schatz, Bildsten & Cumming, 2003). Due to their particular characteristics these superbursts cannot be explained by unstable nuclear burning of hydrogen and/or helium. They appear to be consistent with unstable carbon burning from a deeper layer of heavier elements previously produced by the H/He burning (e.g., Cumming & Bildsten, 2001; Cumming, 2003, Kuulkers, 2004). The large thickness of the fuel layer means that the ignition conditions are sensitive to the thermal profile of the neutron star, opening a new way to probe neutron star interiors (Brown, 2004). The production of carbon is sensitive to the details of H/He burning, testing our understanding of rp-process H burning, and complementing radioactive ion beam studies of heavy proton rich nuclei (Schatz et al., 2003). So far, about 15 of these superbursts have been discovered from archival data analyses (e.g., in ’t Zand, Cornelisse & Cumming, 2004) and they are thus obviously of great interest for further studies. Indeed, the much stronger energy release, and thus much higher fluence of the superbursts, corresponds to fluxes during their long decay times, which can give very high signal to noise spectra making it possible to detect line features from the surface of the neutron stars.

### 1.1.2 Intermediate long bursts

At a few occasions some type I X-ray bursts showing extended decay times up to a few tens of minutes have been observed. As shown on Fig.1, these unusually long bursts have durations, and therefore energy releases, intermediate between the two above mentioned burst classes. INTEGRAL has brought important contributions to the study of such intermediate long bursts, because half of the dozen registered so far have actually been detected by JEM-X (e.g., Brandt et al., 2004; Molkov et al., 2005; Falanga et al., 2008). The mechanisms driving the intermediate long bursts are not yet fully understood: depending on the composition of the accreted material, they may be explained by either the unstable burning of a large pile of mixed hydrogen and helium, or the ignition of a thick pure helium layer. In the latter case, long duration bursts are expected at very low accretion rates (e.g., Cumming et al., 2006; Peng et al., 2007), in particular from Ultra-Compact X-ray Binary (UCXB) systems containing a hydrogen-poor white dwarf (e.g., in 't Zand et al., 2007). However, some intermediate long bursts have been observed at high accretion rates, implying the burning of some hydrogen, and are thus inconsistent with the theoretical picture. As a matter of fact, the relationship between accretion regime, burst types, and recurrence times of most X-ray burst systems is still an open issue for the current theory. It is a function of accretion rate and composition, depending on the nature of the donor star (its age, size, and composition) and the geometry of the binary system, and of the sedimentation and nuclear ignition processes on the surface of the neutron star.

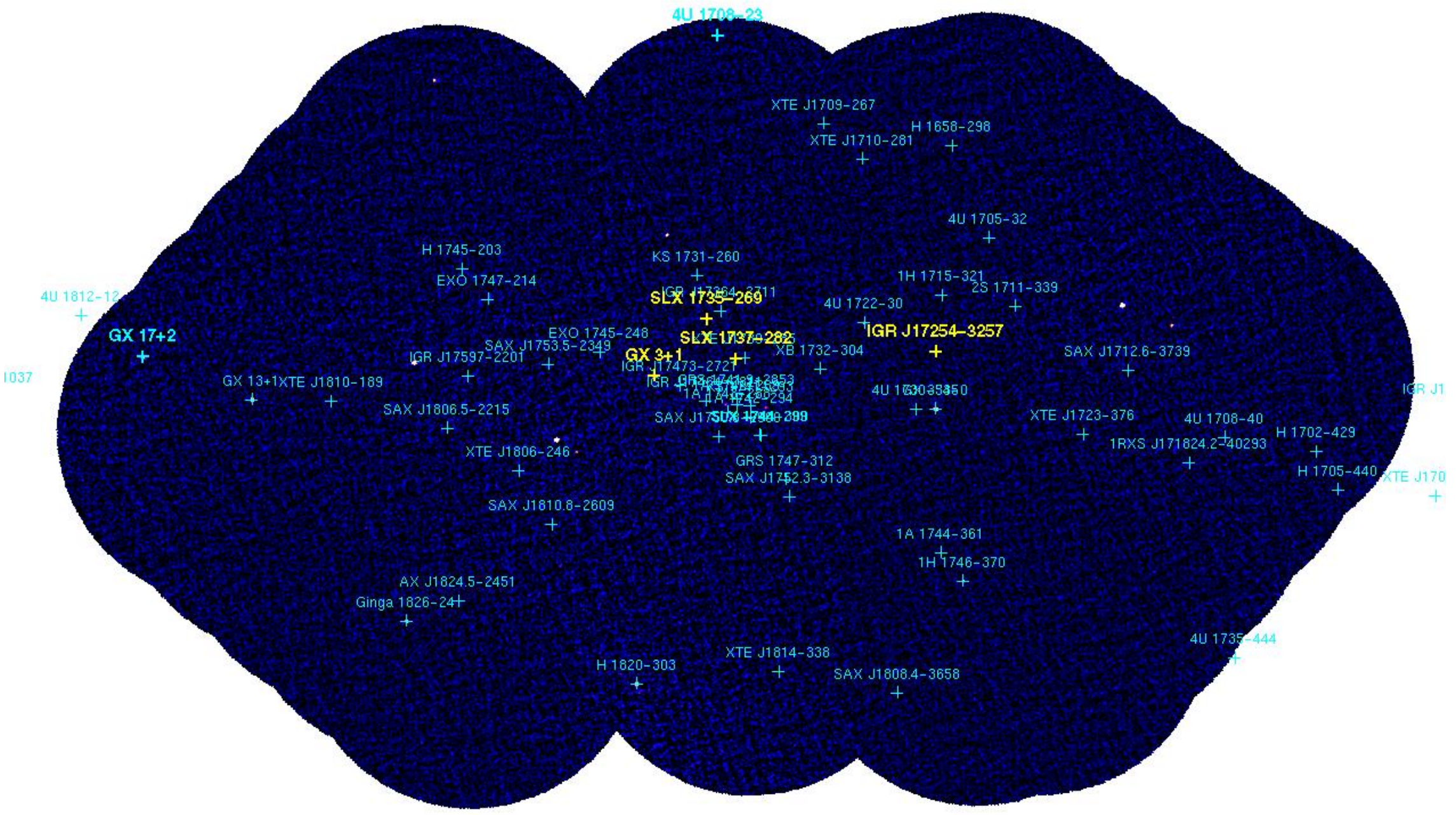

**Figure 2**: JEM-X mosaic image combining several Key Programme observations of the Galactic Centre region and showing the distribution of the X-ray bursters (about 2/3 of the 89 bursters known to date). The highlighted source names indicate the bursters having displayed intermediate long bursts; the yellow ones are the sources whose intermediate long bursts were registered by INTEGRAL.

### 1.2 Monitoring of long X-ray bursts with INTEGRAL

With its exceptional broad band pass INTEGRAL is an important instrument to use to understand more completely the physics of X-ray bursts, specifically the nature of the hard X-ray emission. Moreover, thanks to the unique large field of view of the INTEGRAL instrumentation, it is possible to observe many sources simultaneously, so as to get good exposure for rare events. In particular by covering the Galactic Centre region, where about half of the known X-ray bursters are located (see Fig. 2), the INTEGRAL Key Programmes provide a good opportunity to observe long or super- bursts.

In Table 1, we give the list of the known X-ray bursters (see also in ’t Zand et al., 2004 and references therein) for which we have been given data rights with the aim of continuing the successful series of INTEGRAL detections and analyses of new long bursts. A total of 60 bursters are covered by both IBIS and JEM-X cameras during the INTEGRAL AO-6 Key Programme observations of the Galactic Centre region (48 sources), the Galactic Disk Plane (10 sources), and the Cygnus region (2 sources). Though these sources are the most probable long X-ray burster candidates, this list should only be considered as a natural basis for our monitoring, and we must stay open to the discovery of new X-ray bursters (either actual new bursting sources, or previously known sources but not as X-ray bursters).

Other observation programmes, such as the INTEGRAL Galactic Bulge monitoring (Kuulkers et al., 2007), as well as archive public data, are also exploited in the search of unusual X-ray burst events.

## 2.Detection method and data analysis

As a consequence of their black-body emission at kT ≈ 2 keV, most of the X-ray burst flux is situated in the JEM-X energy range below 10 keV. Therefore a very convenient way to detect X-ray bursts is by inspecting the JEM-X detector light curve recording all good X-ray events in the detector. This method allows us to monitor simultaneously all the sources inside the field of view of the instrument. The search for bursts longer than a few tens of minutes is however not trivial as they can span over more than one single pointing of the satellite (see Fig. 3). Moreover, the detection significance of a burst event depends upon the choice of the light curve time binning, and it may thus be necessary to resample the light curve several times with successively wider time bins to facilitate the detection of a ~½h long burst. When a burst is thus spotted in the detector light curve, it is necessary to compare the image corresponding to the time interval of the burst with an equivalent exposure reference image previous to the burst in order to identify the origin of the event. In most cases[1], it is then convenient to extract the source light curve at the obtained position to get the actual burst count rates corrected for off-axis effects.

Bursts with durations of tens of minutes are actually easier to investigate with the time resolved spectral analysis method than usual short X-ray bursts because of the higher number of accumulated photons. This method consists in extracting the burst spectra during successive time intervals as short as allowed by the counting statistics, making thus possible to follow the spectral evolution of the event. Spectral analysis results obtained by fitting the successive spectra with a black-body model provide the time evolution of the burst flux, as well as the neutron star photosphere radius and temperature.

[1] The extracted source light curve may be contaminated by the emission of other sources inside the field of view. However, in a large majority of situations, an X-ray burst is the predominant source of photons in the instrument. The counts from the burst may be affected only if the bursting source is close to another strong source, or if another burst occurs in the field of view during a long burst.

| | | |
|---|---|---|
| 4U 1708-23 [a] ♦ | SLX 1735-269 ♦ | 4U 1746-37 |
| 4U 1705-32 | XTE J1739-285 | AX J1750.5-2900 |
| XTE J1709 -267 | SLX 1737-282 ♦ | EXO 1747-214 |
| XTE J1710-281 | KS 1741-293 | GRS 1747-312 |
| SAX J1712.6-3739 | GRS 1741.9-2853 | SAX J1752.3-3138 |
| 4U 1711-34 | 1A 1742-289 | SAX J1753.5-2349 |
| 1H 1715-321 | 1A 1742-294 | AX J1754.2-2754 [e] |
| IGR J17191-2821 [b] | GX .2-.2 | IGR J17597-2201 |
| XTE J1723-376 | IGR J17464-2811 | SAX J1806.5-2215 |
| IGR J17254-3257 ♦ | IGR J17473-2721 [d] | XTE J1806-246 |
| 4U 1722-30 | SLX 1744-299 | SAX J1808.4-3658 |
| GX 354-0 | SLX 1744-300 | XTE J1810-189 [f] |
| MXB 1730-335 | GX 3+1 * ♦ | SAX J1810.8-2609 |
| KS 1731-260 * | 1A 1744-361 | XTE J1814-338 |
| XB 1732-304 | 4U 1745-203 | 4U 1820-303 * |
| IGR J17364-2711 [c] | EXO 1745-248 | AX J1824.5-2451 |
| H 1608-522 * | H 1705-440 | 3A 1850-087 |
| 4U 1636-536 * | 4U 1812-12 | 3A 1905+000 |
| XTE J1701-462 [g] | GX 17+2 * ♦ | Cyg X-2 |
| H 1702-429 | Aql X-1 | 4U 2129+47 |

**Table 1**: List[2] of the 60 bursters covered by all INTEGRAL instruments during the Key Programme observations (48 sources in the Galactic Centre region, 10 in the Galactic Disk Plane, and 2 in the Cygnus region). * indicates a known superburster, and ♦ indicates a source having shown at least one intermediate long burst. [a] Hoffman et al. (1978); [b] Klein-Wolt et al. (2007); [c] Chelovekov et al. (2006); [d] Del Monte et al. (2008); [e] Chelovekov & Grebenev (2007); [f] Markwardt et al. (2008); [g] Homan et al. (2007).

[2] Not listed here: XTE J1701-407 is a new X-ray source that showed a long intermediate burst observed by Swift in July 2008 (Falanga et al., 2008; Linares et al., 2008).

## 3. Observation results

During the AO-4 Galactic Centre Key Programme, we observed a 15 min long burst from IGR J17254-3257 (see Fig. 3), which is a source with faint persistent emission and from which only one short burst was previously recorded (Brandt, Budtz-Jørgensen & Chenevez, 2006). Though the long burst just occurred at the end of a stable pointing and continued a piece of time during the next pointing, it was possible to reconstruct the light curve of the event during the slew interval by a simple linear interpolation of the detector event list. The analysis of the two burst observations allowed us to explain the occurrence of both a short and a long bursts from IGR J17254-3257 by a slightly variation of the accretion rate and a burning regime intermediate between pure He and mixed H/He. The long intermediate burst was thus consistent with the burning of a thick layer of helium slowly accumulated by the steady hydrogen fusion into helium, which eventually ignited due to a weak hydrogen flash (Chenevez et al., 2007).

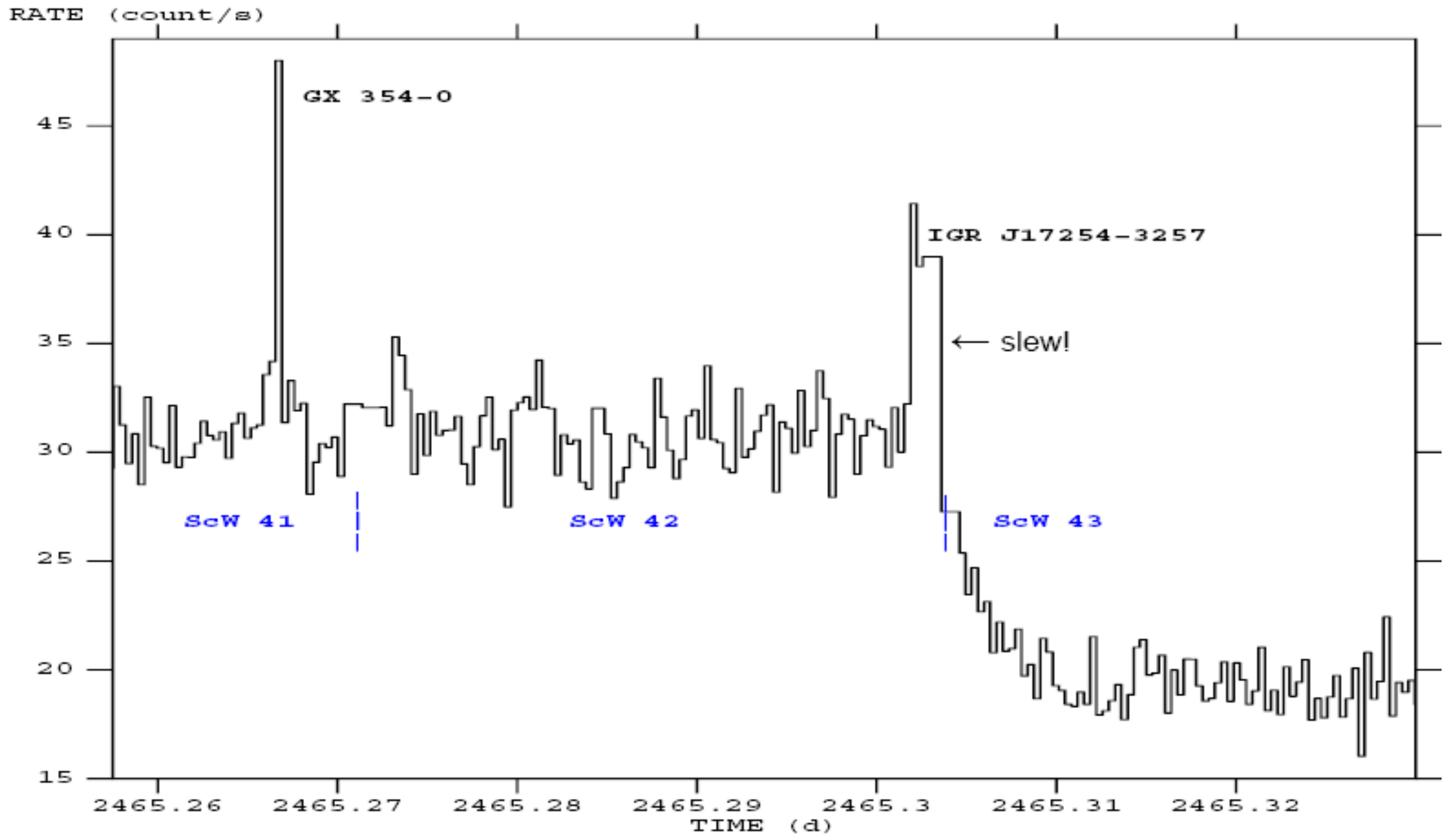

**Figure 3**: Portion of the JEM-X (3-10 keV) detector light curve (time bins of 30 s) obtained during the Key Programme observation of INTEGRAL revolution 484 on October 1st, 2006. It shows a first burst in science window (ScW) #41, subsequently identified as originating from the source GX 354-0, and a second burst from IGR J17254-3257 occurring at the end of ScW #42 and continuing through the satellite slew into ScW #43.

A handful of intermediate long bursts have shown, before the extended decay phase, an initial spike similar to a normal short X-ray flash (see, e.g., Kuulkers et al., 2002). Figure 4 displays the light curve of the two-phase long burst from GX 3+1 observed with INTEGRAL in 2004. The most probable explanation for this burst is the unstable burning of a mixed hydrogen/helium layer involving an unusually large amount of hydrogen. However, we note that this intermediate long burst did occur approximately at the same accretion level as the superburst from the same source reported by Kuulkers (2002). Therefore, it may not be totally ruled out that such peculiar long bursts might be a kind of link between short and super- bursts, where the premature ignition of the carbon layer may be triggered by the helium flash detonation (Chenevez et al., 2006).

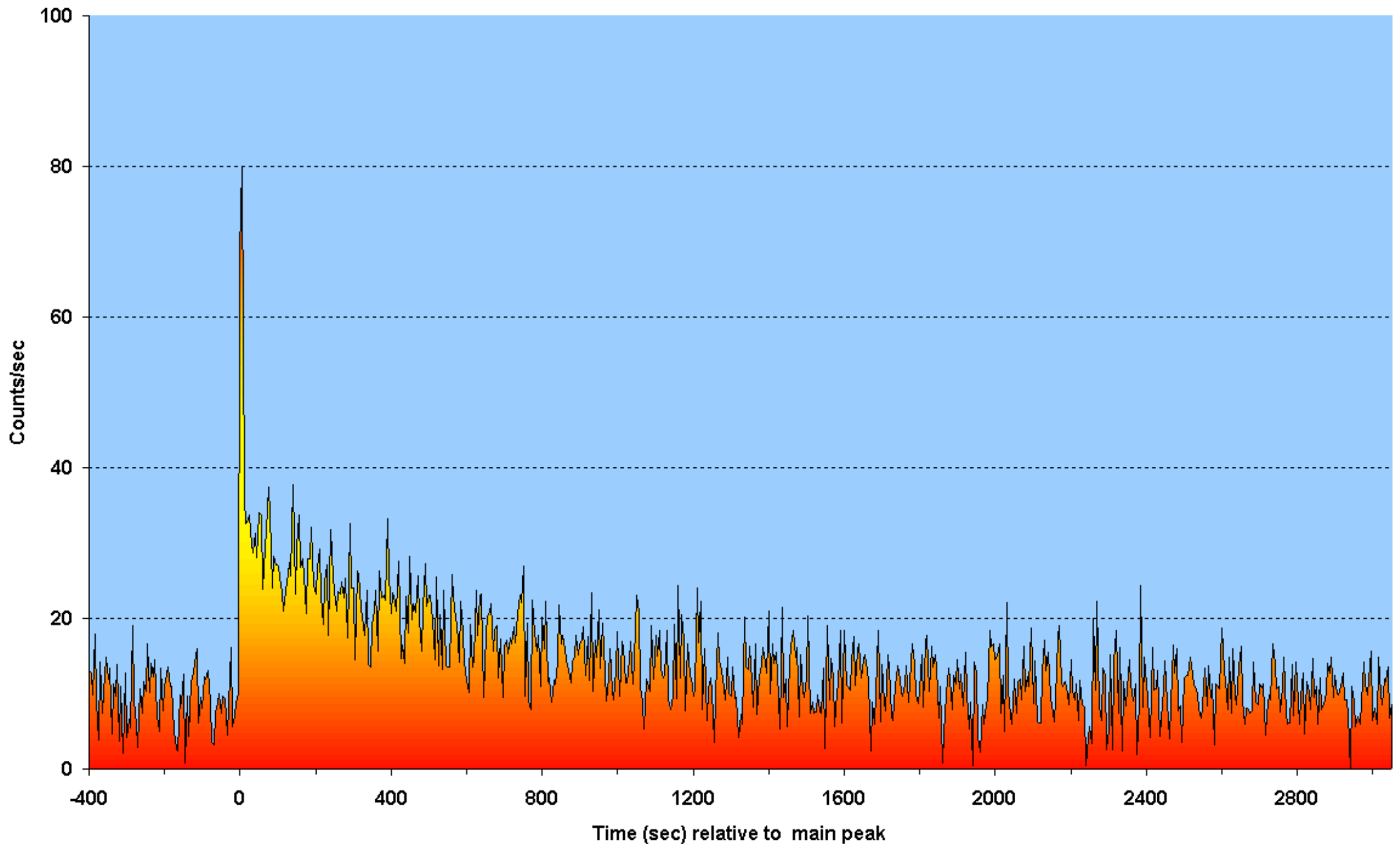

**Figure 4**: JEM-X light curve of the long X-ray burst from GX 3+1 observed by INTEGRAL on August 31, 2004. The energy range is 3-20 keV and the time binning is 5s (INTEGRAL Picture of the Month, February 2006, http://integral.esac.esa.int/POMFeb2006.html).

Of particular interest is the case of the weak persistent source SLX 1737-282 as the only burster that exclusively shows intermediate long X-ray bursts. The first event was observed by BeppoSAX in March 2002 (in 't Zand et al., 2002) and three more similar events were observed by INTEGRAL in March 2004, April 2005, and April 2007, respectively (Falanga et al., 2008). Figure 5 shows the JEM-X and ISGRI light curves of these three bursts and Fig. 6 presents the corresponding time resolved spectral analysis results.

The second of the above bursts, which also occurred during two consecutive stable pointings (it was not possible to reconstruct the event during the data gap), presents the interesting characteristics of a photospheric radius expansion of the neutron star, as witnessed by the short flux variation immediately after the peak in the JEM-X and ISGRI light curves. This corresponds to a rapid and large variation of the apparent black-body radius simultaneously with a short decrease of the black-body temperature. This is due to the bolometric luminosity that reaches the Eddington limit for which the outwards radiation pressure overcomes the gravity. For a canonical equation of state of a neutron star of 1.4 solar masses and a 10 km radius, this corresponds to a so-called Eddington luminosity from which it is thus possible to derive the distance to the source. In the case of SLX 1737-282, we obtained a distance of 7.3 kpc.

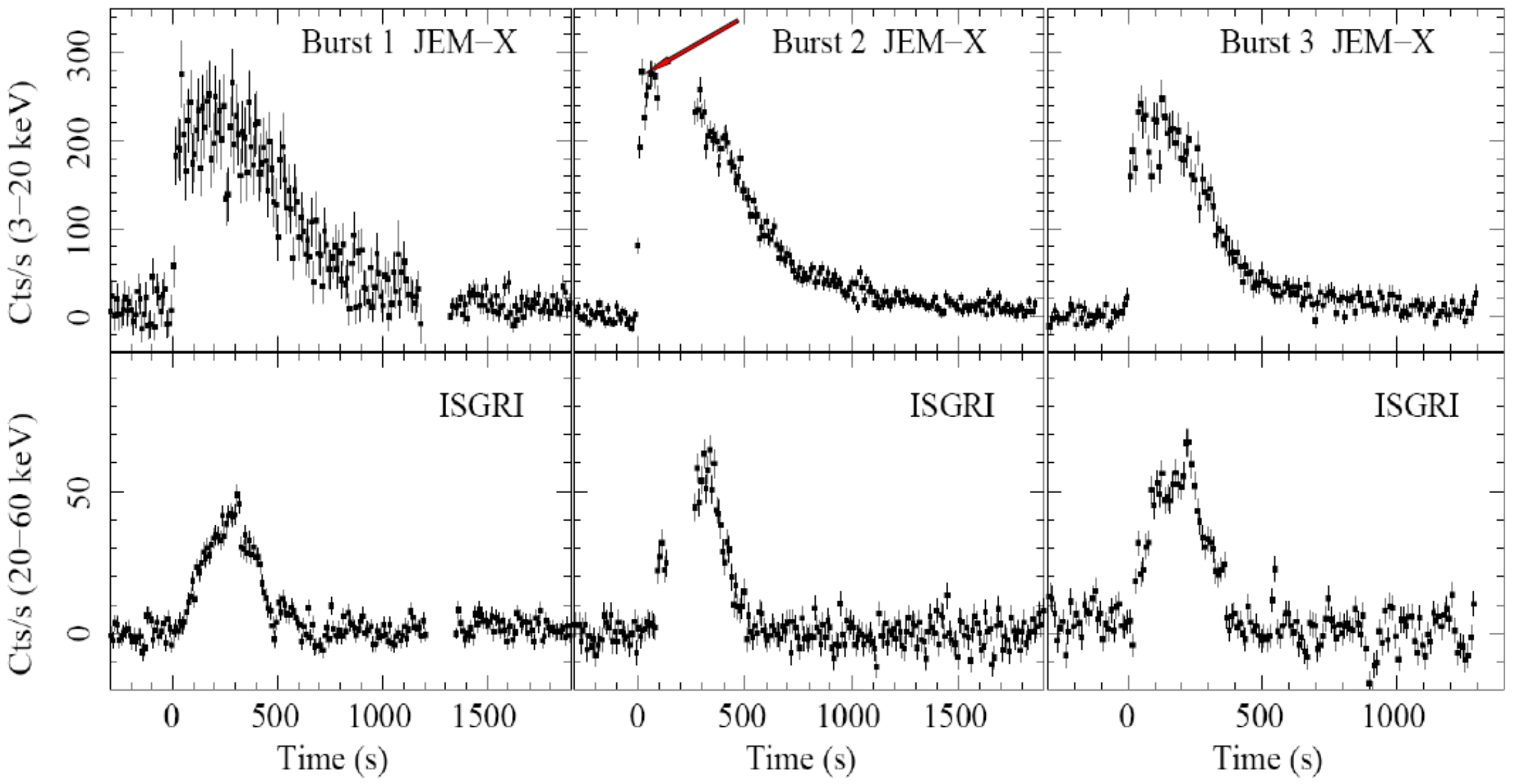

**Figure 5**: JEM-X and ISGRI light curves of the bursts from SLX 1737-282 (Falanga et al., 2008).

The singular bursting behaviour of SLX 1737-282, together with its weak persistent emission, is consistent with a picture where the neutron star is in an UCXB system as suggested by in 't Zand et al. (2007). In that picture, the long bursts are explained by the burning of a thick layer of helium slowly accreted from a pure helium donor, such as a degenerated white dwarf (Falanga et al., 2008).

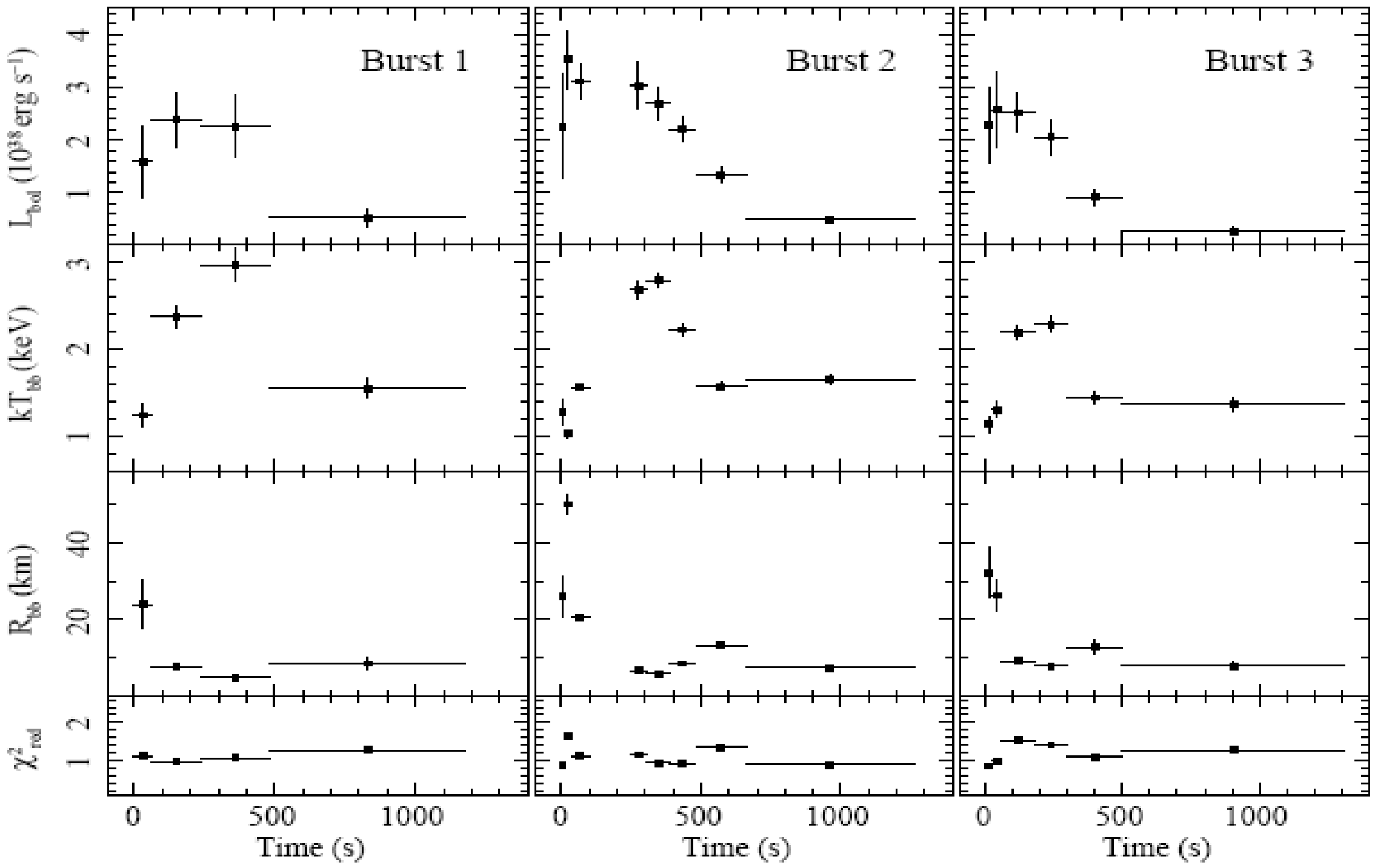

**Figure 6**: Temporal evolution of the black-body fit spectral parameters: bolometric luminosity at 7.3 kpc, corresponding radius, and temperature of the photosphere, as well as the quality of the fits displayed by the $\chi^2_{red}$ values (Falanga et al., 2008).

In Table 2, we summarize the main results of our monitoring of long X-ray bursts with INTEGRAL. To be complete we add the first intermediate long burst observed by INTEGRAL in 2003 from the source SLX 1735-269 (Molkov et al., 2005). At that time, these authors did explain this burst by the unstable burning of a large pile of mixed hydrogen and helium on the surface of the neutron star. More recent research, as presented here, allows now us to interpret the long burst from SLX 1735-269 in the same way as for SLX 1737-282. Indeed, considering the relative low accretion rate of this source, we can conclude that this burst is consistent with the unstable burning of a large pile of pure helium, likely slowly accreted from a pure helium donor in an UCXB system, as also suggested by in ‘t Zand et al. (2007).

| Source | Date | $T_b$ (s) \| $\tau$ (s) | $E_b$ (erg) | Acc. Rate ($g/cm^2/s$) | Burning regime | Reference |
|---|---|---|---|---|---|---|
| GX 3+1 | 20040831 | 1800 \| 131 | $2 \cdot 10^{40}$ | 10 000 | He → H | Chenevez et al., 2006 |
| IGR J17254-3257 | 20061001 | 900 \| 216 | $2 \cdot 10^{40}$ | 400 | H → He | Chenevez et al., 2007 |
| SLX 1737-282 | 20040309 | 1500 \| 275 | $0.7 \cdot 10^{41}$ | 800 | He | Falanga, Chenevez et al., 2008 |
| | 20050411 | 1800 \| 323 | $1.2 \cdot 10^{41}$ | | He | |
| | 20070402 | ~900 \| 281 | $1.0 \cdot 10^{41}$ | | He | |
| SLX 1735-269 | 20030915 | 2000 \| 400 | $2 \cdot 10^{41}$ | 1 500 | He | Molkov et al., 2005 |

**Table 2**: Summarized results for each of the six intermediate long X-ray bursts observed by INTEGRAL/JEM-X. $T_b$ and $\tau$ are respectively the approximate total burst duration and the exponential decay. $E_b$ is the total energy release. The burning regime indicates the sequence and/or the main fuel (see text).

## 4. Future prospects

Thermonuclear bursts from accreting neutron stars have been investigated for many years. One scientific objective is to resolve the complexity of the relationship between nuclear burning and accretion process regimes, in order to interpret the diversity of the X-ray burst observations. A few bursting sources, which have been studied intensively, offer confirmed examples of three classes of ignition predicted theoretically as a function of the accretion rate (Fujimoto et al., 1987). However, the mechanisms driving the long bursts, in particular, are not yet fully understood and have recently been the subject of advanced theoretical research (e.g., Cumming et al. 2006; Peng et al. 2007).

Looking forward the hopeful observation of the first superburst by INTEGRAL, the long term goal of the present work is to interpret the various types of thermonuclear bursts into a consistent picture of the ignition and burning processes in relation with the accretion regime of the neutron stars. An improved investigation of these various processes, based on the study of a large number of events, will further our knowledge about the composition of the companion stars, the geometry of the accretion flow onto the neutron stars, and therefore the evolution of low-mass X-ray binaries, as well as a better understanding of the nuclear burning physics. It is thus necessary to acquire larger data sets in order to strengthen the statistics of global studies. In that respect, it is necessary to develop the search for X-ray bursts in JEM-X detector light curves as an automated process.

The fluxes reached during an X-ray burst can give spectra with high signal-to-noise ratio, making it possible to detect atomic line features from the photosphere of the neutron stars. Such detection would therefore enable a measurement of the gravitational redshift of the neutron star (e.g., Cottam, Paerels, and Mendez, 2002). Moreover, the analysis of X-ray bursts in the gamma-ray energy range may provide a unique opportunity to detect nuclear spectral lines from such objects. A significant detection would imply that radioactive material is brought, during the burst, to the upper layers of the atmosphere, or even ejected from the neutron star. Indeed, photospheric radius expansion bursts likely eject nuclear burning ashes that are potentially detectable as photoionisation edges (Weinberg et al. 2006). Furthermore, the strong energy release of long and super- bursts may yield much larger amounts of ejected material than for ordinary radius expansion bursts, which may ease the detection of the emitted lines.

Moreover, a primary objective for the INTEGRAL mission is the detection and study of gamma-ray emission related to nucleosynthesis. Therefore, the INTEGRAL instrumentation offers for the first time an opportunity to search for nuclear gamma-rays produced by nucleosynthesis during the X-ray bursts. A positive detection of gamma-rays from X-ray bursters would thus probe the thermonuclear reactions and open a new window into the secluded stellar element synthesis processes.

In conclusion, it is worth continuing to monitor long X-ray burst events by taking advantage of the unique instrumentation of INTEGRAL.

## Acknowledgements

This research was supported by ESA-PRODEX contract Nº 90057.